\documentclass[conference]{IEEEtran}
\usepackage{amsmath,amssymb,amsthm}    % need for subequations
\usepackage[dvips]{graphicx}   % need for figures
\usepackage{verbatim}   % useful for program listings
\usepackage{color}      % use if color is used in text
\usepackage{subfigure}  % use for side-by-side figures
\usepackage{epstopdf}
\usepackage{float}%\usepackage{keywords}
\usepackage{algorithm}
\usepackage{algorithmic}
\usepackage{setspace}
\usepackage{subfigure}
\usepackage{cite}
\DeclareMathOperator*{\argmin}{arg\,min}
\makeatletter
\newcommand*{\rom}[1]{\expandafter\@slowromancap\romannumeral #1@}
\makeatother

\begin{document}
\title{\textbf{Joint Sparsity Pattern Recovery with 1-bit Compressive Sensing in Sensor Networks}}

%\name{Vipul Gupta$^\textit{1}$, Bhavya Kailkhura$^\textit{2}$, Thakshila Wimalajeewa$^\textit{2}$ and Pramod K.
%Varshney$^\textit{2}$}
%\address{}
\author{\authorblockN{{Vipul Gupta$^1$, Bhavya Kailkhura$^2$, Thakshila Wimalajeewa$^2$, and Pramod K. Varshney$^2$}}\\
\authorblockA{$^{1}$Indian Institute of Technology Kanpur, Kanpur 208016, India\\
$^{2}$Department of EECS, Syracuse University, Syracuse, NY 13244, USA}}
\maketitle

\begin{abstract}
We study the problem of jointly sparse support recovery with 1-bit compressive measurements in a sensor network.   
Sensors are assumed to observe sparse signals having the same but unknown sparse support.
Each sensor quantizes its measurement vector element-wise to 1-bit and transmits the quantized observations to a fusion center.
We develop a computationally tractable support recovery algorithm which minimizes a cost function defined in terms of the likelihood function and the $l_{1,\infty}$ norm.  
We observe that even with noisy 1-bit measurements, jointly sparse support can be recovered accurately with multiple sensors each collecting only a small number of measurements. 
\end{abstract}

\footnotetext[1]{This work is  supported in part by the National Science
Foundation (NSF) under Grant No. 1307775.}


\section{Introduction}
Support recovery of a sparse signal deals with the problem of finding the locations of the non-zero elements of the sparse signal. This problem occurs in a wide variety of areas including source localization \cite{Malioutov1,Cevher1},
sparse approximation \cite{Natarajan1}, subset selection in linear regression \cite{Miller1,Larsson1}, estimation of frequency band locations in cognitive radio networks \cite{Tian1},
and signal denoising \cite{SSChen1}. In these applications, finding the
support of the sparse signal is more important than recovering the complete signal itself. The problem of sparsity pattern recovery has been addressed by many authors in the last decade in
different contexts. \emph{Compressive sensing (CS)} has recently been introduced as a sparse signal acquisition scheme via random projections. 
%With CS, the problem of support recovery of sparse signals has received much attention in the context of random dictionaries. 
A good amount of work has already been done for support recovery with real valued measurements \cite{wain2,Fletcher1,Akcakaya1, tang1}. 
However, in practice, measurements are quantized before transmission or storage, therefore, it is important to consider quantization of compressive measurements for practical purposes.
%It is important to consider quantization of compressive measurements since in practice, measurements are quantized before transmission or storage. 
Further, coarse quantization of measurements  is desirable and/or  even necessary in resource constrained communication networks.  There are some recent
works that have addressed the problem of recovering
sparse signals/sparsity pattern  based on quantized
compressive measurements in different
contexts,  be it calculating performance bounds \cite{jstsp12,note_on_support_recovery} or devising recovery algorithms \cite{1bit_supp_recovery,boyd,1bit_lp,baraniuk1}.

%Use of 1-bit CS  is particularly attractive because of the low communication bandwidth it requires for data transfer. 
However, most of the work on 1-bit CS has focused only on recovery for the single sensor case. Reliable recovery of a sparse signal based on 1-bit CS is very difficult with only one sensor, especially when the signal-to-noise ratio (SNR) is low. % A fair amount of work has already been done on quantized compressive sensing for the single sensor case, for example, \cite{microsoft,1bit_supp_recovery,boyd,baraniuk1,1bit_lp}.
On the other hand, simultaneous  recovery of  multiple sparse signals  arises naturally in a number of applications including distributed sensor and cognitive radio networks.
%Some algorithms have been provided for signal recovery using 1-bit measurements  in \cite{baraniuk1,microsoft,1bit_lp,1bit_supp_recovery} with only a single measurement vector.
To the best of our knowledge, the problem of jointly sparse support recovery with multiple sensors based on 1-bit CS has not been explored in the literature.  In this work, we exploit the benefits of using multiple nodes for jointly sparse recovery with 1-bit CS measurements. We assume that the multiple nodes observe sparse signals with the same but unknown  sparsity pattern. The measurement vectors at each node are quantized to 1-bit element-wise and transmitted to a fusion center.  

To recover the jointly sparse support, we propose to solve an optimization problem which minimizes an objective function expressed in terms of the likelihood function and the $l_{1,\infty}$ norm of a matrix. We use a computationally tractable algorithm to recover the common sparsity pattern.
%We formulate an optimization problem which minimizes an objective function expressed in terms of likelihood function and $l_{1,\infty}$ norm of a matrix,  and use greedy algorithms to recover the common sparsity pattern.
 %from the sign bits of the compressed signal.
 We show that by employing multiple sensors, the common sparse support can be estimated reliably with a relatively small number of 1-bit CS measurements per node. 
 In particular, we investigate the trade-off between the possibility of deploying multiple sensor nodes and the cost of sampling per node. 

\section{Observation Model}
We consider a distributed network with multiple nodes that observe sparse signals having the same sparse support.  Let the number of sensors be $P$. At a given node, consider the following $M \times 1$ real valued observation vector collected via random projections:
 \begin{equation}\label{observation}
 \mathbf{y}_p = \boldsymbol{\Phi}_p\mathbf{s}_p + \mathbf{v}_p
 \end{equation}
 \\
where $\boldsymbol{\Phi}_p$ is the $M \times N$ ($M<N$) measurement matrix at the $p$-th node for $ p = 1, \cdots ,P$ and $N$ is the signal dimension. For each $p$, the entries of $\boldsymbol{\Phi}_p$  are assumed to be drawn from a Gaussian ensemble with mean zero. The sparse signal vector of interest, $\mathbf{s}_p$ for $p=1,\cdots,P$, has only $K (\leq N)$ nonzero elements with the same support. The measurement noise vector, $\mathbf{v}_p$, at the $p$-th node,  is assumed to be independent and identically distributed (i.i.d.) Gaussian with mean vector $\mathbf 0$ and covariance matrix $\sigma_v^2 \mathbf I_M$
where $\mathbf 0$ is a vector of all zeros   and   $\mathbf I_M$ is the $M\times M$ identity matrix.

Let each element  of $\mathbf y_p$ be  quantized to 1-bit  so that the $i$th  quantized measurement at the $p$th node is given by,

%Let $\mathbf Y$ be a $M\times P$ matrix in which the $p$-th column is $\mathbf y_p$ for $p =1,2,\cdots,P$. Similarly, let $\mathbf{S}$ be the $N \times P$ matrix which contains $\mathbf{s}_p$ as its columns for $p = 1,2,\cdots,P$. In the following, we assume that the signal matrix $\mathbf{Y}$ is quantized element-wise into one of two levels which requires 1-bit per element of $\mathbf Y$:

\begin{eqnarray}
z_{ip}=\left\{
\begin{array}{ccc}
0, ~&~ \textrm{if}~ -\infty < y_{ip} < 0\\
1, ~&\textrm{if}~ 0 \leq y_{ip} < \infty,\\
\end{array}\right.\label{quant1}
\end{eqnarray}
\\
where $y_{ip}$ is the $i${th} element of $\mathbf y_p$, for $i=1,2,\cdots,M$ and $p=1, \cdots, P$. 
%In particular, we take sign measurements of the signal and, therefore, zero acts as the quantizer threshold.
Let $\mathbf Z$  and $\mathbf Y$ be   $M\times P$ matrices in which the $(i,p)$-th element of $\mathbf Z$ and $\mathbf Y$ are $z_{ip}$ and $y_{ip}$ respectively, for $p = 1,2,\cdots,P$ and $i=1,\cdots, M$. Further, let   $\mathbf{S}$ be the $N \times P$ matrix which contains $\mathbf{s}_p$ as its columns for $p = 1,2,\cdots,P$.  In matrix notation, (\ref{quant1}) can be written as, 
\begin{eqnarray}
\mathbf Z = \mathrm{sign}(\mathbf Y)\label{quant2}
\end{eqnarray}
where $\mathrm{sign}(\mathbf Y)$ denotes the sign of each element of $\mathbf Z$.

%Let $\mathbf{R}$ be the matrix obtained after quantizing the compressed signal $\mathbf{Y}$ (which is also an $M \times P$ matrix).
%$$\mathbf{R} = \begin{bmatrix} r_{11} & \cdots & r_{1P} \\ \vdots & \ddots & \vdots \\ r_{m1} & \cdots & r_{mP} \end{bmatrix}$$
%where
%\begin{eqnarray}\label{observation_2}
%r_{ip} = z_{ip} + w_{ip},
%\end{eqnarray}
% for $i = 1,2,....,M$  and $p = 1,2,....,P;$	
%and  $w_{ip}$ is the decoder noise which is assumed to be iid Gaussian
%with mean zero and variance $\sigma_w^2$.

\section{Common Support Recovery with 1-bit CS Measurements via $\l_1$-Regularized Maximum Likelihood}\label{algo}
%\subsection{$l_1$-Regularised Maximum Likelihood}

%\begin{figure}
%\includegraphics[width=3.6in, height=2.3in]{percent_varyP_m_50_100.eps}
%\caption{Support Recovery Performance with Varying Number of Sensors (P) with m constant for each curve}
%\end{figure}
%
%\begin{figure}
%\includegraphics[width=3.6in, height=2.3in]{prob_varyP_m_50_100.eps}
%\caption{Probability of recovering exact Support with Varying Number of Sensors (P) with m constant for each curve}
%\end{figure}

In this section, first we formulate an optimization problem for joint sparsity pattern recovery for 1-bit CS.
We use the regularized $l_1$ norm  minimization approach with  the likelihood function as the cost function instead of the widely used least squares function. With quantized measurements, the approach which uses the likelihood function as the cost function has been shown to provide better results compared to least squares methods with a single sensor \cite{boyd}.

For the sake of tractability, we assume that the measurement matrix $\boldsymbol \Phi_p= \boldsymbol \Phi$ is the same for all $p = 1,2,\cdots,P$\footnote{The work can easily be extended to the scenario having different measurement matrices.}. We have from (\ref{observation}),

\begin{equation}
y_{ip} = \boldsymbol \Phi_i^T \mathbf{s}_p + v_{ip},
\end{equation}
for $ i = 1,2,\cdots, M$ and $ p = 1,2,\cdots, P$. In the rest of the paper, $\boldsymbol \Phi_i$ denotes  the $i$-th row of $\boldsymbol \Phi$.

%As given by (\ref{quant2}),  $\mathbf Z$  contains  element-wise sign measurements in (\ref{quant1}).  
Next, we calculate probabilities $\Pr(z_{ip} =1)$ and $\Pr(z_{ip} =0)$ which will later be used to write the expression for the likelihood of $\mathbf{Z}$ given $\textbf{S}$. We have,
\begin{equation*}
\Pr(y_{ip} \geq 0)
\Rightarrow \Pr(\boldsymbol \Phi_i^T \mathbf{s}_p + v_{ip} \geq 0)=\phi (\boldsymbol \Phi_i^T \mathbf{s}_p/\sigma_v).
\end{equation*}

Similarly,
\begin{equation*}
\Pr(y_{ip} < 0)
\Rightarrow \Pr(\boldsymbol \Phi_i^T \mathbf{s}_p + v_{ip} < 0)= \phi (-\boldsymbol \Phi_i^T \mathbf{s}_p/\sigma_v).
\end{equation*}
where $\phi(x) = (1/\sqrt{2\pi})\int_{-\infty}^xe^{-t^2/2}dt$.
The conditional probability of $\mathbf Z$ given $\mathbf S$ is given by,
\begin{eqnarray*}
&~&\Pr (\textbf{Z}|\textbf{S}) = \prod_{p=1}^P \prod_{i=1}^M \Pr (z_{ip}|\textbf{S})\\
&=& \prod_{p=1}^P \prod_{i=1}^M \left(\phi \left(\frac{\boldsymbol \Phi_i^T \mathbf{s}_p}{\sigma_v}\right)\right)^{z_{ip}} \times \left(\phi\left(- \frac{ \boldsymbol \Phi_i^T \mathbf{s}_p}{\sigma_v}\right)\right)^{(1-z_{ip})}.
\end{eqnarray*}

The negative log-likelihood of $\mathbf{Z}$ given $\textbf{S}$, $f_{ml}(\boldsymbol\Phi \mathbf S)$, is given by

\begin{footnotesize}
$$-\sum_{p=1}^P \sum_{i=1}^M \left[z_{ip}\log\left(\phi \left(\frac{\boldsymbol \Phi_i^T \mathbf{s}_p}{\sigma_v}\right)\right) + (1-z_{ip})\log\left(\phi\left(- \frac{ \boldsymbol \Phi_i^T \mathbf{s}_p}{\sigma_v}\right)\right)\right]$$
\end{footnotesize}
which can be rewritten as

\begin{multline}\label{f_ml}
f_{ml}(\textbf{X}) = -\sum_{p=1}^P\sum_{i=1}^m \left[z_{ip}\log\left(\phi\left(\frac{x_{ip}}{\sigma_v}\right)\right)\right. \\
+ \left.(1-z_{ip})\log\left(\phi\left(\frac{-x_{ip}}{\sigma_v}\right)\right)\right],
\end{multline}
and $\textbf{X} = \boldsymbol \Phi \textbf{S}$. In the following, we use $\mathbf X$ and $\Phi \mathbf S$ interchangeably.
We need to minimize this expression, $f_{ml}(\boldsymbol{\Phi}\mathbf{S})$,  as well as incorporate the sparsity condition of the signal matrix $\mathbf{S}$ to obtain an estimated signal matrix $\hat{\mathbf{S}}$ or the support of $\mathbf{S}$. As all the signals observed at  all the nodes have the same support, the row-$l_0$ norm (as defined in \cite{trop} for real valued measurements) is appropriate to incorporate the joint sparsity constraint. 
%Simulations also complement the definition of the problem in the manner shown below with good results.
The row-$l_0$ norm of $\mathbf S$ is given by,
$$||\mathbf{S}||_{row-0} = |\textrm{rowsupp}(\mathbf{S})|,$$ which is also referred to as the $l_{0,\infty}$ norm, where the
row support of the coefficient matrix $\mathbf{S}$ is defined as \cite{trop}
$$\textrm{rowsupp}(\mathbf{S}) = \{w \in [1,N]: s_{wk} \neq 0 ~\textrm{for some}~ k\}.$$
Now to compute $\mathbf S$, one can solve the following optimization problem:
\begin{equation}\label{prob2}
\underset{\mathbf{S}}{\arg\min}~\{f_{ml}(\boldsymbol{\Phi}\textbf{S})+ \lambda||\mathbf{S}||_{0,\infty}\}
\end{equation}
where $\lambda$ is the penalty parameter.
However, the problem (\ref{prob2}) is not tractable in its current form and can be relaxed as
%\cite{trop}
\begin{equation}\label{main_prob1}
\underset{{\mathbf{S}}}{\arg\min}~\{f_{ml}(\boldsymbol{\Phi}\textbf{S})+ \lambda||\mathbf{S}||_{1,\infty}\}
\end{equation}
where $||\mathbf{S}||_{1,\infty} = \sum_{i=1}^N \underset{1 \leq j \leq P}{\max} |s_{ij}|$, i.e., $||\mathbf{S}||_{1,\infty}$ is the sum of all the elements with maximum absolute value in each row, also known as the $l_{1,\infty}$ norm of a matrix.  

The goal is to 
develop a computationally tractable algorithm to
solve the problem of the form
\begin{equation} \label{main_prob}
\underset{\mathbf{S}}{\arg\min}~\{f(\boldsymbol{\Phi}\textbf{S})+ \lambda g(\mathbf{S})\}
\end{equation}
where $f(\boldsymbol{\Phi}\textbf{S}) = f_{ml}(\boldsymbol{\Phi}\textbf{S})$ and $g(\mathbf{S})$ is the $l_{1,\infty}$ norm of $\mathbf{S}$.

%\subsection{Algorithm for Solving the Optimization Problem}
%\vspace{-2mm}
%\emph{Applying ISTA with constant step-size}:
We use \emph{iterative shrinkage-thresholding} algorithms (ISTA) for solving the problem defined in (\ref{main_prob}). In ISTA, each iteration involves solving a simplified optimization problem, which in most of the cases can be easily solved using the proximal gradient method, followed by a shrinkage/soft-threshold step; for e.g., see \cite{ista3,ista6,FISTA}.
From \cite{FISTA}, at the $k$-th iteration we have
\begin{equation}{\label{update_xk}}
\mathbf{S}_k = P_{L_f}(\mathbf{S}_{k-1})
\end{equation}
where
\begin{equation}{\label{pl_xk-1}}
P_{L_f}(\mathbf{T}) = \underset{\hat{\mathbf{S}}}{\argmin}~\lambda g(\mathbf{S}) + \frac{L_f}{2}||\mathbf{S} - (\mathbf{T} - \frac{1}{L_f}\nabla f(\mathbf{T})||_F^2.
\end{equation}

Inputs to the algorithm are $L_f$ (the Lipshitz constant of $\nabla f$) and $\mathbf{S}_0$, the initialization for the iterative method, which can be kept  null matrix or $\boldsymbol{\Phi}^{\dagger}\mathbf{Z}$, where $\boldsymbol{\Phi}^{\dagger}$ is the pseudoinverse of $\boldsymbol{\Phi}$ and $\mathbf{Z}$ is the quantized received signal matrix as defined before.
For our case, the gradient of $f_{ml}(\mathbf X)$ w.r.t. matrix \textbf{S} can be easily calculated as $\boldsymbol \Phi^T \nabla f_{ml}(\textbf{X})$, where $\mathbf X= \boldsymbol \Phi \textbf{S}$. Notice that, $\nabla f_{ml}(\textbf{X})$ is the gradient of $f_{ml}(\textbf{X})$ w.r.t. \textbf{X} and is given by

\begin{equation}\label{gradml}
\nabla f_{ml}(x_{ip}) = \frac{z_{ip}\textrm{exp}(-\frac{\tilde{x}_{ip}^2}{2})}{\sqrt{2\pi}\sigma_v\phi (\tilde{x}_{ip})} -
\frac{(1-z_{ip})\textrm{exp}(-\frac{\tilde{x}_{ip}^2}{2})}{\sqrt{2\pi}\sigma_v\phi (-\tilde{x}_{ip})},
\end{equation}
where $\tilde{x}_{ip} = x_{ip}/\sigma_v.$

%However, we cannot solve (\ref{pl_xk-1}) directly. We rather divide the problem into various subproblems and then solve each of the subproblems separately.
The problem defined in (\ref{pl_xk-1}) is row separable for each iteration. Therefore, to solve for $\mathbf{S}_k$, i.e., to find $P_L(\mathbf{S}_{k-1})$, we divide the problem into $N$ subproblems, where $N$ is the number of rows in $\mathbf{S}$. Next, we solve the following subproblem for each row of $\mathbf{S}_k$:
\begin{equation}{\label{subproblem}}
\underset{\mathbf{s}^i}{\argmin}~\lambda g(\mathbf{s}^i) + \frac{L_f}{2}||\mathbf{s}^i - (\mathbf{t}^i - \frac{1}{L_f}\nabla f(\mathbf{t}^i))||_2^2;
\end{equation}
where $\mathbf{s}^i$, $\mathbf{t}^i$ and $\nabla f(\mathbf{t}^i)$ are the $i^{th}$ rows of $\mathbf{S}$, $\mathbf{S}_{k-1}$ and $\nabla f(\mathbf{S}_{k-1})$ respectively. Equation (\ref{subproblem}) is of the form:
\begin{equation}\label{sub_prob1}
\underset{\mathbf{s}}{\arg\min}~\left\{\lambda g(\mathbf{s}) + \frac{L_f}{2}||\mathbf{s} - \mathbf{u}||_2^2\right\};
\end{equation}
where  $g(\mathbf{s}) = ||\mathbf{s}^i||_\infty,$ i.e., the $l_\infty$ norm of the $i^{th}$ row of $\mathbf{S}$  and constant vector $\mathbf{u}$ is given by
$\mathbf{u} = \mathbf{t}^i - \frac{1}{L_f}\nabla f(\mathbf{t}^i)$ (we do not use superscript $i$ on $g(\mathbf s)$ and $\mathbf u$ for brevity).

For (\ref{sub_prob1}), we have the following equivalent problem in the epigraph form 
\begin{equation}\label{op_problem}
\underset{\mathbf{s}}{\argmin}~ \left\{\bar{\lambda}t + \frac{1}{2}||\mathbf{s} - \mathbf{u}||_2^2\right\},~ \textrm{s.t.}~ 0 \leq \text{sgn}(u_p)s_p \leq t,
\end{equation}
where
$\bar{\lambda} = \frac{\lambda}{L_f}$,
$u_p$ and $s_p$ are the $p$-th elements in $\mathbf u$ and $\mathbf s$ respectively, for all $p = 1,2, \cdots, P$ and $t\in \mathbb{R}$. 
Define $w_p = \text{sgn}(u_p)$. The problem in (\ref{op_problem}) can be solved using Lagrangian based methods. The Lagrangian for (\ref{op_problem}) is 
$$L(\mathbf{s},t,\mathbf{\alpha},\mathbf{\beta}) = \bar\lambda t + \frac{1}{2}\|\mathbf{s}-\mathbf{u}\|_2^2 + \sum_p \beta_p(w_ps_p - t) - \sum_p\alpha_pw_ps_p$$
with dual variables $\mathbf \alpha = [\alpha_1,\cdots, \alpha_P]^T$ and $\mathbf \beta = [\beta_1,\cdots, \beta_P]^T$. 

Hence, for strong duality to hold, following Karush–-Kuhn–-Tucker (KKT) conditions must be satisfied by the optimal $\mathbf{s}^*,t^*,\mathbf{\alpha}^*$ and $\mathbf{\beta}^*$.
\begin{eqnarray}\label{op_coditions}
&&(s_p^* - u_p) - \alpha_p^*w_p + \beta_p^*w_p = 0, \label{eq1} \\
&&\bar{\lambda} - \sum_p \beta_p^* = 0, \label{eq2} \\
&&\alpha_p^*(w_ps_p^*) = 0, \label{eq3}\\
&&\beta_p^*(w_ps_p^* - t^*) = 0,\label{eq4}\\
&&\alpha_p^*,\beta_p^* \geq 0.\label{eq5}
\end{eqnarray}

Note that $0 \leq w_ps_p \leq t, \; p = 1, \ldots, P$. To find the optimal $\mathbf{s}^*$, consider three simple cases \\

\noindent\textit{Case (i)}: 
\begin{eqnarray*}
&& w_ps_p^* = 0 \\
&\Rightarrow& \beta_p^* = 0 \; \text{(from (\ref{eq4}))}\\
&\Rightarrow& u_p + \alpha_p^*w_p = 0 \; \text{(from (\ref{eq1}))}\\
&\Rightarrow& \alpha_p^* = \frac{-u_p}{w_p} \; \text{which is} \leq 0.
\end{eqnarray*}

Therefore, from (\ref{eq5}), $s_p^* = 0$ if and only if $u_p=0$.\\

\noindent\textit{Case (ii)}: 
\begin{eqnarray*}
&& 0<w_ps_p^* <t\\
&\Rightarrow& \beta_p^* = 0 \; \text{(from (\ref{eq4})) and}\; \alpha_p^* = 0 \; \text{(from (\ref{eq3}))}\\
&\Rightarrow& s_p = u_p\; \text{(from (\ref{eq1}))} 
\end{eqnarray*}

\noindent\textit{Case (iii)}: 
\begin{eqnarray*}
&& w_ps_p^* =t\\
&\Rightarrow& s_p - u_p + \beta_p^*w_p = 0 \;\text{ (from (\ref{eq1}))}\\
&\Rightarrow& \frac{t^*}{w_p} - u_p + \beta_p^*w_p = 0 \\
&\Rightarrow& \beta_p^* = \frac{u_p}{w_p} - \frac{t^*}{w_p^2} \\
&\Rightarrow& \beta_p^* = |u_p| - t^*. 
%(w_p = \text{sgn}(u_p), \text{thus},\; \frac{u_p}{w_p} = |u_p|\;\text{ and}\; w_p^2 = 1). 
\end{eqnarray*}
Also, since $\beta_p \geq 0 ~\forall~ p = [1,2,\cdots,P]$, we have $\beta_p^* = (|u_p| - t^*)_+$, where $(x)_+$ is defined as max$(x,0)$.\\
Using (\ref{eq2}) in the above equation, we have 
\begin{equation}\label{bisection}
\sum_p(|u_p| - t^*)_+ - \bar\lambda = 0
\end{equation}
as $\beta_p = (|u_p| - t^*)_+$.
This can be easily solved for $t^*$ by applying the bisection based method using the initial interval $[0,||u||_\infty]$. Define 
$$g(t) = \sum_p(|u_p| - t^*)_+ - \bar\lambda.$$ 
Therefore, $g(t^*) = 0$.
If there exists no solution in the interval $[0,||u||_\infty]$, i.e., $g(0) \times g(||u||_\infty) \geq 0$, the trivial solution is given by $t^* = 0.$
Once we have the optimal $t^*$, the optimal $s^*$ is given by
\begin{equation}\label{solution}
s_p^* = \left\{
\begin{array}{c}w_pt^* ~\textrm{if}~ |u_p| \geq t^*;\\
u_p ~ \textrm{otherwise}.\\
\end{array}\right.
\end{equation}

Each subproblem given by (\ref{subproblem}) can be solved in a similar way, the solution to each of which can be used to find $\mathbf{S}_k$ using (\ref{update_xk}) and (\ref{pl_xk-1}). The summary of all the steps is provided in Algorithm \ref{algo1} where $||.||_F$ denotes the Frobenius norm.
Algorithm \ref{algo1} produces the matrix $\mathbf S_k$ and locations of non-zero elements in $\mathbf{S}_k$ gives the support of original signal matrix $\mathbf S$.

\begin{algorithm}
%\setstretch{0.5}
\begin{enumerate}
%\vspace{-1mm}
\item \textbf{Given} tolerance $\epsilon {>} 0$, parameters $ \tilde{\lambda} {>} \lambda,~ 0{<}\alpha {<} 1 ~\textrm{and} ~L_f$
%\vspace{-6mm}
\item \textbf{Initialize} $\mathbf {S}_0$ ($\mathbf S_{-1} = \mathbf S_0$),  $ ~\hat {\lambda} = \tilde{\lambda}$, $k=0$
%\vspace{-2mm}
\item
\textbf{While} $~\hat {\lambda} > \lambda$
%\vspace{-2mm}
\item
$\hat {\lambda} = \alpha\hat{\lambda}$
%\vspace{-2mm}
\item
\textbf{While} $||\mathbf{S}_k - \mathbf{S}_{k-1}||_{F} > \epsilon ||\mathbf{S}_{k-1}||_{F}$
%\vspace{-2mm}
\item
$k = k+1$
%\vspace{-2mm}
\item
%for all $i \in \{1,2,\cdots,P\}:$
Define matrix $\mathbf{U} = \mathbf{S}_{k-1} - \frac{1}{L_f}\nabla f(\mathbf{S}_{k-1})$ where $\nabla f(\mathbf{S}_{k-1})$ is computed as in (\ref{gradml})
%\vspace{-2mm}
\item
\textbf{For} each row of $\mathbf S_k$
\item Update the $p$-th row element using (\ref{solution}) for $p = 1,2,\cdots,P$
%\vspace{-6mm}
\item
\textbf{End For}
%\vspace{-2mm}
\item
\textbf{End While}
%\vspace{-2mm}
\item
\textbf{End While}\label{algo_1}
%\vspace{-2mm}
\end{enumerate}
 \caption{Estimation of the common support of the sparse signal
 }\label{algo1}
 \end{algorithm}
 
  \begin{figure*}[htb]
 \begin{minipage}[b]{.48\linewidth}
  \centering
  \centerline{\includegraphics[width=8.0cm]{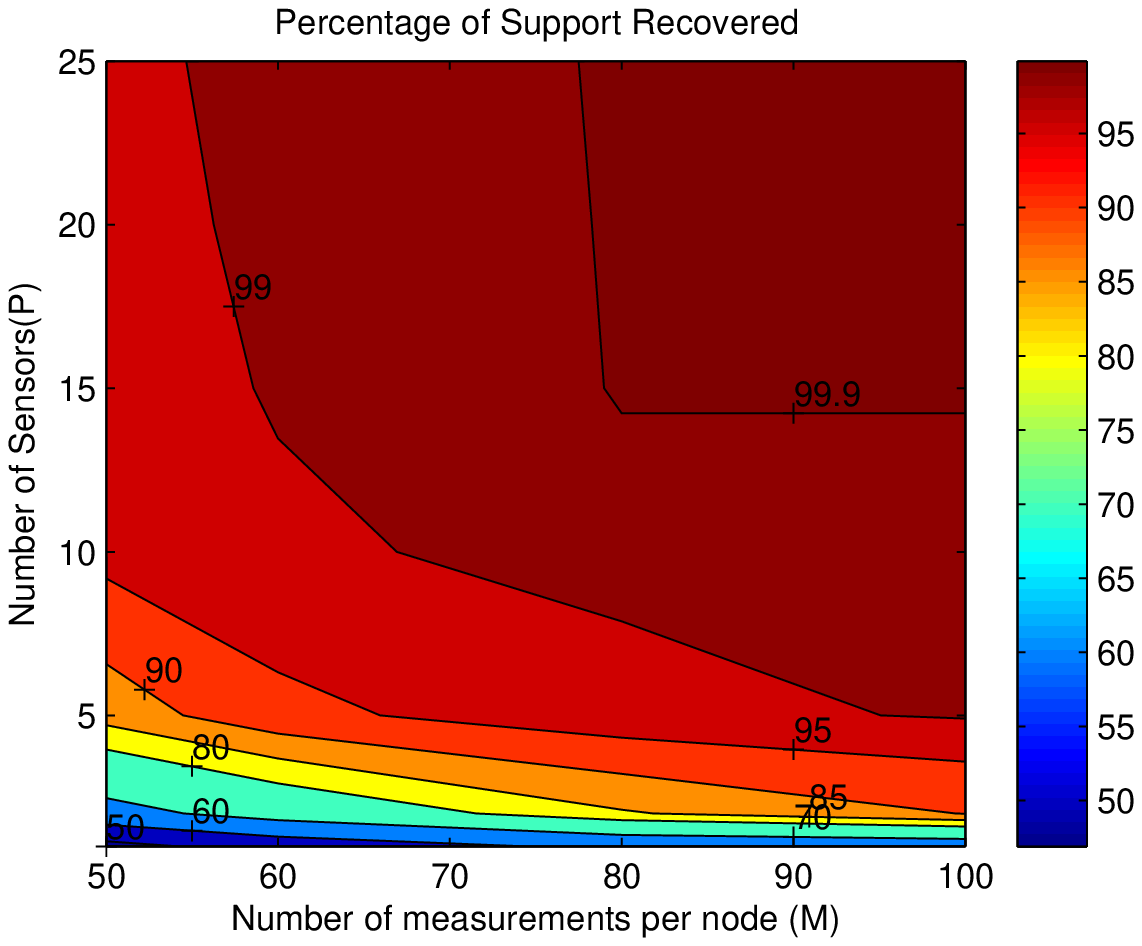}}
%  \vspace{1.5cm}
  \centerline{(a) Percentage of Support Recovered}\medskip
\end{minipage}
\hfill
\begin{minipage}[b]{0.48\linewidth}
  \centering
  \centerline{\includegraphics[width=8.0cm]{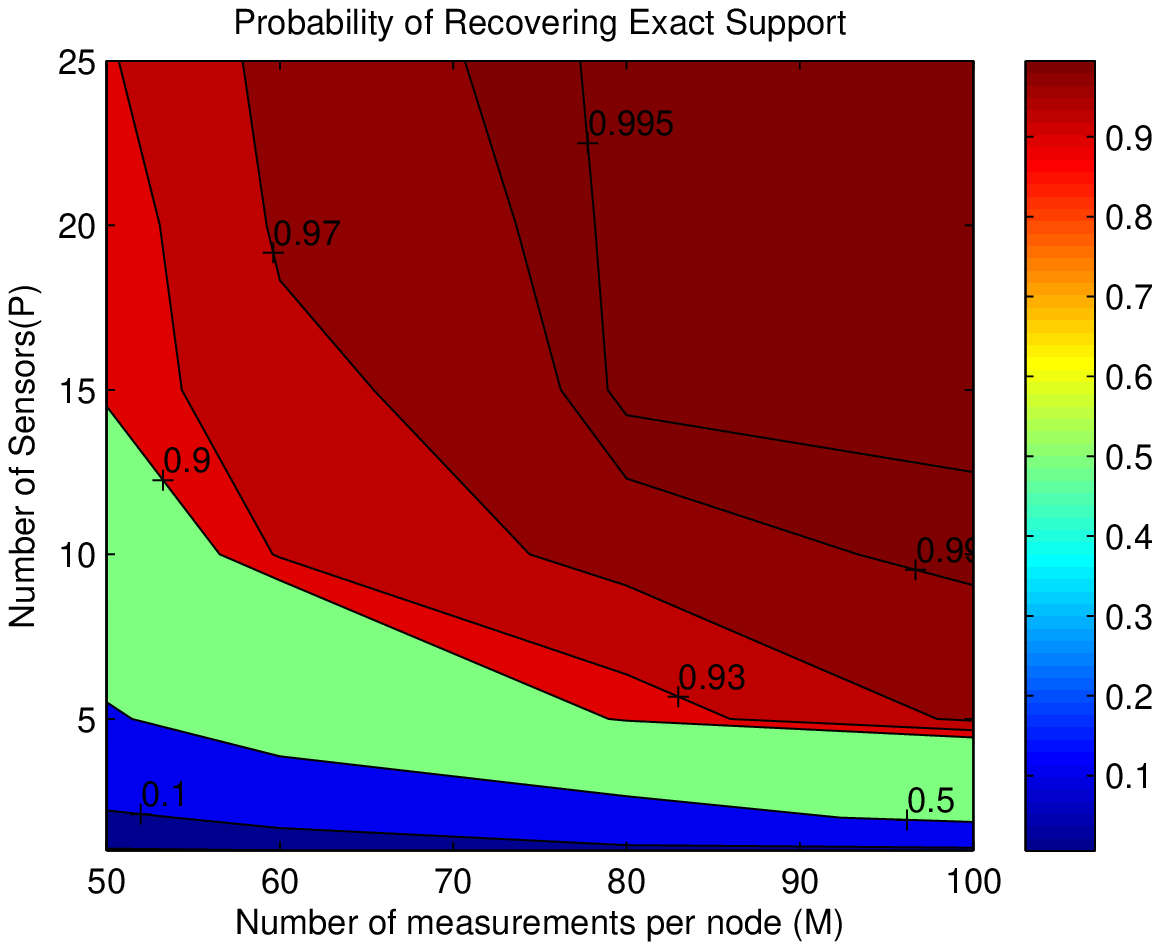}}
%  \vspace{1.5cm}
  \centerline{(b) Probability of Recovering Exact Support }\medskip
\end{minipage}
\caption{Performance of common sparsity pattern recovery when $\sigma_v^2=0.01$ }
\label{fig_Low_SNR}
\end{figure*}

 \begin{figure*}[htb]
 \begin{minipage}[b]{.48\linewidth}
  \centering
  \centerline{\includegraphics[width=8.0cm]{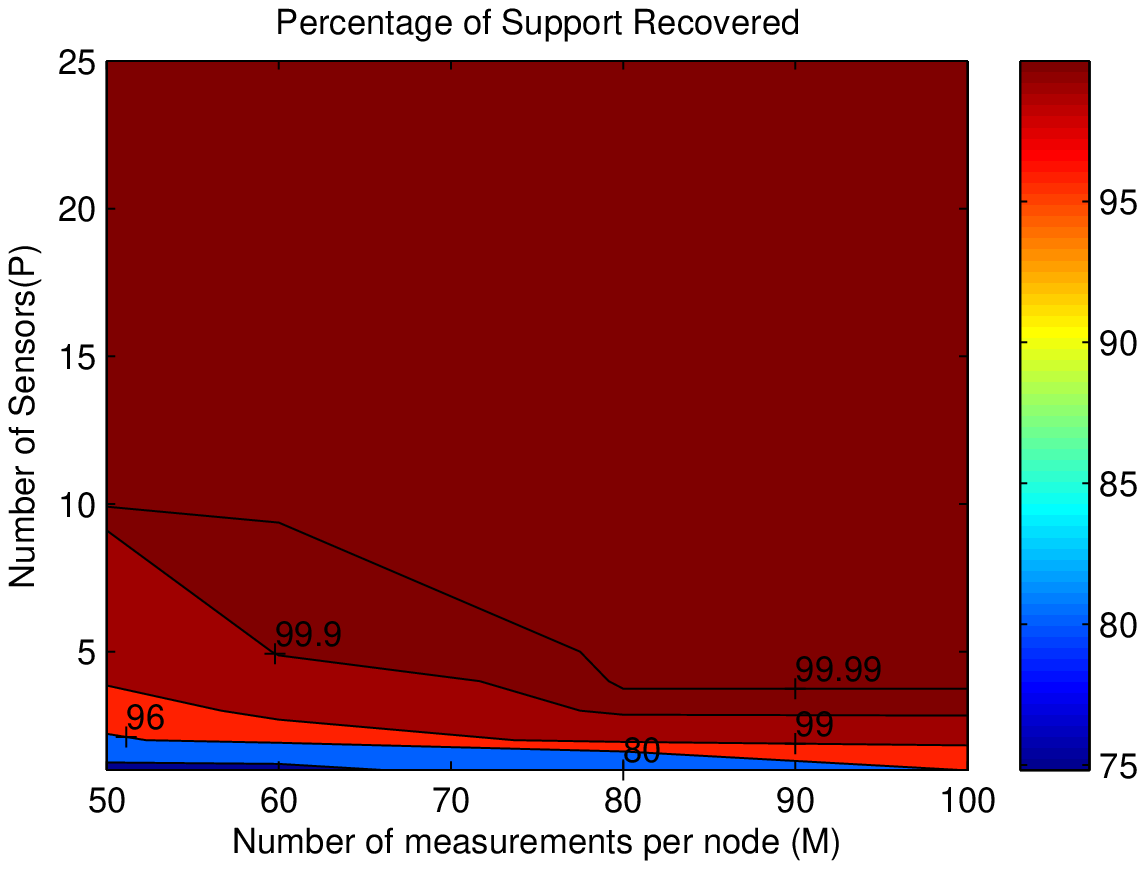}}
%  \vspace{1.5cm}
  \centerline{(a) Percentage of Support Recovered}\medskip
\end{minipage}
\hfill
\begin{minipage}[b]{0.48\linewidth}
  \centering
  \centerline{\includegraphics[width=8.0cm]{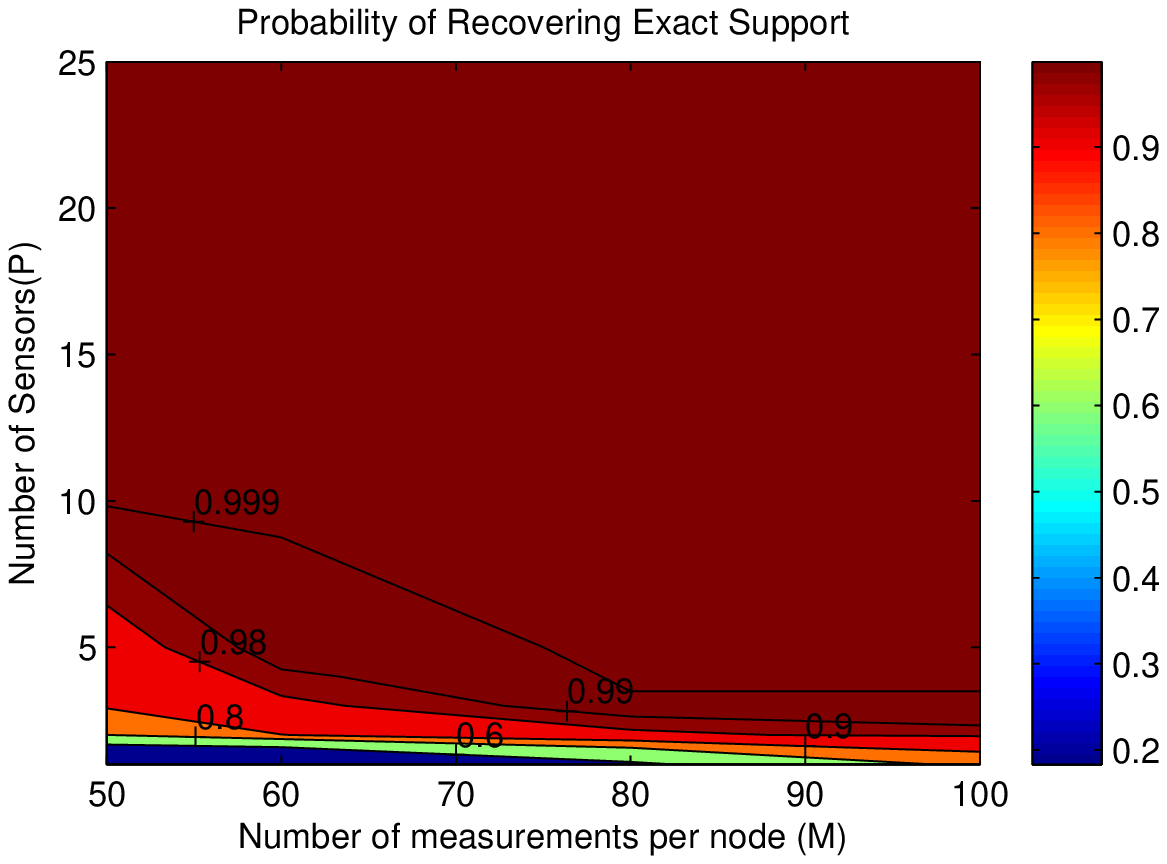}}
%  \vspace{1.5cm}
  \centerline{(b) Probability of Recovering Exact Support }\medskip
\end{minipage}
\caption{Performance of common sparsity pattern recovery when $\sigma_v^2=0.0001$ }
\label{fig_High_SNR}
\end{figure*}

 \begin{figure*}[htb]\label{fig_comparison}
 \begin{minipage}[b]{.48\linewidth}
  \centering
  \centerline{\includegraphics[width=7.0cm]{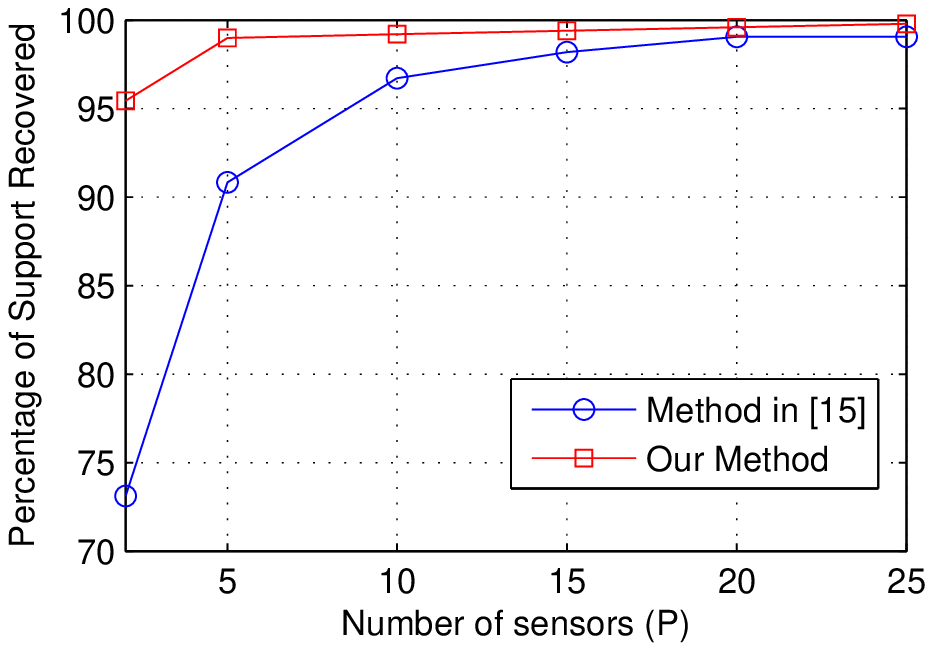}}
%  \vspace{1.5cm}
  \centerline{(a) Percentage Support Recovery}\medskip
\end{minipage}
\hfill
\begin{minipage}[b]{0.48\linewidth}
  \centering
  \centerline{\includegraphics[width=7.0cm]{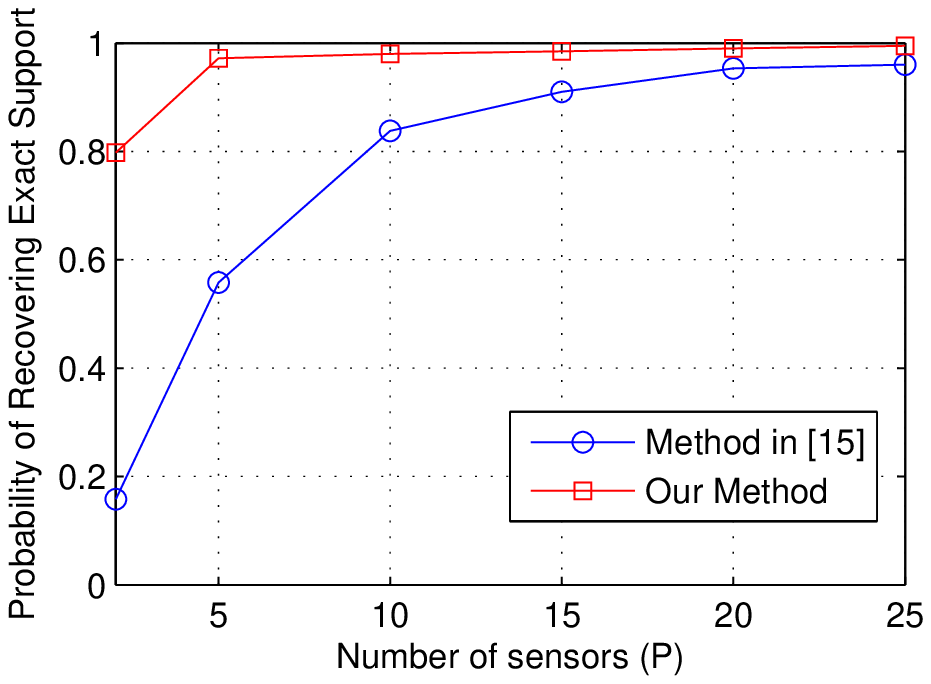}}
%  \vspace{1.5cm}
  \centerline{(b) Probability of Estimating Exact Support}\medskip
\end{minipage}
\caption{Comparison of our results with the approach presented in \cite{boyd}: $M=50$, $N=100$ and $K=5$, in terms of the percentage of support recovered correctly  and the probability of recovering exact support }
\label{fig_High_SNR2}
\end{figure*}

\section{Numerical Results}
In this section, we present some simulation results to demonstrate the performance of jointly sparse support recovery with 1-bit CS based on   our proposed algorithm. For every Monte Carlo run, we generate the elements of the $M \times N$ measurement matrix $\boldsymbol \Phi$ from a normal distribution with mean zero and variance = 0.004. In all our simulations, we used randomly generated sparse signal matrix $\mathbf{S}$, with each column size fixed as $N = 100$. 
%The number of columns is given by the number of sensors $P$. 
We choose $K$ random values out of $N$, which are the nonzero rows in $\mathbf S$. For each column of $\mathbf{S}$, all the elements whose position is given by the $K$ random values are assigned a value of either $1$ or $-1$ with probability $0.5$. The value of $K$ is kept $5$. The observation noise is assumed to be Gaussian with mean zero and variance is $\sigma_{v}^2 = 0.01$ (SNR = 2.96 dB, low SNR case), and $\sigma_{v}^2 = 0.0001$ (SNR = 23.01 dB, high SNR case).  We measure the percentage of support recovered correctly and the probability of recovering exact support as  $M$, $P$ and SNR vary using $1000$ Monte Carlo runs.

Our results for low SNR regime are plotted in Fig. \ref{fig_Low_SNR}. The y-axis shows the number of sensors ($P$) and the x-axis shows the number of measurements per node ($M$). In  Fig. \ref{fig_Low_SNR} (a), the numbers on contours represent the percentage of support that are recovered. Similarly, in Fig. \ref{fig_Low_SNR}(b), the numbers on contours represent the probability of recovering the exact support. We can deduce that for a particular value of $M$, the performance improves with the number of sensors, and vice-versa. Similarly, Fig. \ref{fig_High_SNR} shows  the contour plots in the high SNR regime. We see that even with a very small number of sensors, for e.g., $P = 3$, the algorithm performs very well for reasonable values for $M$, such as $M = 60$.

In Fig. \ref{fig_High_SNR2}, we compare our results with one of the most related algorithms for sparse recovery with quantized measurements as provided in \cite{boyd}, which uses the ML method for only one node (one measurement vector). To compute the common support with $P$ measurement vectors, the individual support sets were computed  for each signal using the algorithm in \cite{boyd}, and the estimated support sets were fused using the majority rule.
For comparison, values of $M$, $N$ and $K$ were chosen as 50, 100 and 5, respectively. The observation noise variance $\sigma_v^2=0.0001$. As seen from Fig. \ref{fig_High_SNR2}, our proposed approach for joint sparsity pattern recovery with quantized measurements outperforms the case when  the support is estimated individually as in \cite{boyd} and then fused. In particular, the proposed algorithm in this paper exploits  the jointly sparse nature of multiple measurement vectors thus outperforming the results obtained by estimating the supports individually using \cite{boyd} and then fused.

%providing better results than estimating the supports individually and then fused.

\section{Conclusion}

In this paper, we exploited the use of multiple sensors for the  recovery of  common sparsity pattern of  sparse signals  with  1-bit CS.  A computationally tractable algorithm was developed to  optimize  an objective function defined in terms of likelihood function and the $l_{1,\infty}$ norm of a matrix.   Numerical results show that, with very coarsely  quantized measurements  (only the sign information),   the  common sparsity pattern of sparse  signals   can be recovered reliably  even in the low SNR region, and the performance increases monotonically with the number of sensors.

\bibliographystyle{IEEEtran}
\bibliography{IEEEabrv,bib2}

\end{document}